\documentclass[prd,amsmath,amssymb,superscriptaddress,preprintnumbers,twocolumn,nofootinbib,10pt]{revtex4-1}
\usepackage{graphicx}
\usepackage{dcolumn}
\usepackage{bm}
\usepackage{amssymb}
\usepackage{latexsym}
\usepackage{booktabs}
\usepackage{amsmath}
\usepackage{multirow}
\usepackage{url}
\usepackage{footnote}
\usepackage{float}
\usepackage{threeparttable}
\usepackage[colorlinks=true, linkcolor=blue, citecolor=blue]{hyperref}
\usepackage[bottom]{footmisc}

\usepackage[normalem]{ulem}
\usepackage{color}
\usepackage{array}
\usepackage{enumerate}
\usepackage{adjustbox}

\usepackage{makecell}
\usepackage{diagbox}
\usepackage{epstopdf}
\usepackage{epsfig}
\usepackage{longtable}
\usepackage{supertabular}
\usepackage{algorithm}
\usepackage{pifont}
\usepackage{algorithmic}
\usepackage{changepage}
\usepackage{setspace}

\begin{document}

\title {Examining a new form of non-standard dark matter using DESI DR2 data}

\author{Yan-Hong Yao}\thanks{Corresponding author}
\email{yaoyanhong@nbu.edu.cn}
\author{Yi-Hao Shen}
\affiliation{Institute of Fundamental Physics and Quantum Technology, Department of Physics, School of Physical Science and Technology, Ningbo University, Ningbo, Zhejiang 315211, China} 
\author{Tian-Nuo Li}
\author{Guo-Hong Du}
\affiliation{Liaoning Key Laboratory of Cosmology and Astrophysics, College of Sciences, Northeastern University, Shenyang 110819, China}
\author{Yungui Gong}\thanks{Corresponding author}
\email{gongyungui@nbu.edu.cn}
\affiliation{Institute of Fundamental Physics and Quantum Technology, Department of Physics, School of Physical Science and Technology, Ningbo University, Ningbo, Zhejiang 315211, China} 
\begin{abstract}
In this work, we propose a non-standard dark matter (NSDM) model in which the equation of state (EoS) of dark matter (DM) is parameterized as $w_{\rm dm} = w_2 a^2$, and this DM model is motivated by the idea that DM must become cold dark matter (CDM) in the neighborhood of the scale factor $a = 0$, which implies that both the EoS of DM, $w_{\rm dm}$, and its derivative with respect to the scale factor, ${\rm d}w_{\rm dm}/{\rm d}a$, vanish at $a = 0$. By incorporating the latest cosmological datasets — including the Planck2018 Cosmic Microwave Background (CMB) distance priors, the Baryon Acoustic Oscillation measurements from the Data Release 2 of the Dark Energy Spectroscopic Instrument (DESI), together with three independent Type Ia Supernova datasets, namely the Dark Energy Survey Year 5 (DESY5) compilation, the Union3 compilation, and the PantheonPlus sample — we constrain the $\Lambda w_2$DM, $ww_2$DM, and $w_0w_aw_2$DM models, which are constructed by replacing CDM with NSDM in the $\Lambda $CDM, $w$CDM, and $w_0w_a$CDM models, respectively. We find that there is a preference for a negative DM EoS at more than the $3\sigma$ confidence level for the data combinations CMB+DESI+Union3 and CMB+DESI+DESY5. Moreover, for all data combinations, replacing CDM with NSDM in the $w$CDM and $w_0w_a$CDM models significantly reduces the probability of violating the null energy condition. Furthermore, both $ww_2$DM and $w_0w_aw_2$DM are favored over $\Lambda $CDM with a significance comparable to that of the $w_0w_a$CDM model.
\textbf{}
\end{abstract}
\maketitle

\section{Introduction}
\label{intro}
Dark matter (DM), as a mysterious component of the universe, has its existence supported by multiple astronomical observations, such as galaxy rotation curves~\cite{rubin1970rotation}, gravitational lensing~\cite{tyson1990detection,clowe2004weak,massey2007dark}, the cosmic microwave background (CMB)~\cite{smoot1992structure,spergel2003first,spergel2007three,hinshaw2013nine,ade2014planck,planck2016planck,aghanim2020planck-1}, and large-scale structure~\cite{davis1985evolution,baugh1993three,colless20012df,percival20012df,tegmark2004three,eisenstein2005detection}. From these astronomical observations, several characteristics of DM can be inferred: it interacts only very weakly with other particles aside from gravity, remains non-relativistic, and is extremely stable, hardly undergoing any decay. Moreover, it constitutes a considerable proportion of the universe’s total energy density. Based on these properties, physicists have proposed several candidates for DM, namely Weakly Interacting Massive Particles, axions, and neutralinos, which are collectively referred to as cold dark matter (CDM) candidates. In cosmological studies, these CDM candidates can be represented within a unified phenomenological framework — as an ideal fluid that interacts with other cosmic components only through gravity, and has a zero equation of state (EoS) parameter, zero sound speed, and zero viscosity sound speed. This phenomenological CDM component, together with a cosmological constant that plays the role of dark energy (DE), forms the well-known standard cosmological model — the $\Lambda$CDM model, which fits a wide range of cosmological observations across various scales~\cite{riess1998observational,perlmutter1999measurements,spergel2003first,spergel2007three,hinshaw2013nine,ade2014planck,planck2016planck,aghanim2020planck-1,eisenstein2005detection,anderson2014clustering,alam2017clustering,alam2021completed}.

Despite the success of this paradigm, the CDM hypothesis still requires further testing against more observational data. In fact, some small-scale observations have already revealed phenomena that may hint at deviations from the CDM hypothesis, such as the missing satellite problem~\cite{klypin1999missing,moore1999dark}, the too-big-to-fail problem~\cite{boylan2012milky}, and the core-cusp issue~\cite{moore1999cold,springel2008aquarius}. These persistent small-scale discrepancies have led to the development of alternative DM candidates beyond CDM, including warm DM~\citep{blumenthal1982galaxy,bode2001halo}, fuzzy DM~\citep{hu2000fuzzy,marsh2014model}, interacting DM~\citep{spergel2000observational}, and decaying DM~\citep{wang2014cosmological}. On the other hand, at larger observational scales, researchers have been trying to investigate whether observational data support non-cold DM from the beginning of this century~\cite{mueller2005cosmological,Kumar:2012gr,kumar2014observational,Murgia:2017lwo,Gariazzo:2017pzb,Murgia:2018now,kopp2018dark,Schneider:2018xba,kumar2019testing,ilic2021dark,Najera:2020smt,pan2023iwdm,yao2024observational,Yao:2023qve,Wang:2025zri,Kumar:2025etf,Yang:2025ume,Liu:2025mob,Yao:2025kuz,Li:2025eqh,Li:2025dwz}, although these studies have so far yielded no definitive evidence for non-cold DM, they have provided insightful approaches for modifying the standard cosmological model.

Recently, the Baryon Acoustic Oscillation (BAO) measurements from the Dark Energy Spectroscopic Instrument (DESI) have been released, the combination of DESI BAO data, CMB data, and Type Ia Supernova (SN Ia) data indicates a significant preference for dynamical dark energy (DDE) within the $w_0w_a$CDM model~\cite{DESI:2025zgx}, which have motivated extensive research on various DE models~\cite{Li:2024qso,Giare:2024gpk,Dinda:2024ktd,Escamilla:2024ahl,Sabogal:2024yha,Li:2024qus,Li:2024hrv,Wang:2024dka,Huang:2025som,Li:2025owk,Wu:2025vfs,Li:2025ula,Li:2025ops,Barua:2025ypw,Yashiki:2025loj,Ling:2025lmw,Goswami:2025uih,Yang:2025boq,Pang:2025lvh,You:2025uon,Ozulker:2025ehg,Cheng:2025lod,Pan:2025qwy,Li:2025muv,Du:2025xes}. All of these examined DE models not only deviates significantly from a cosmological constant but also shows a tendency for its EoS parameter to cross $-1$, suggesting a violation of the null energy condition. Some researchers suggest  using non-standard dark matter (NSDM) instead of DDE to alleviate the tensions between the DESI BAO measurements and other cosmological probes to avoid the violation of the null energy condition~\cite{Chen:2025wwn,Giani:2025hhs,Braglia:2025gdo}. For example, Ref.~\cite{Braglia:2025gdo} proposed a NSDM model in which DM is modeled to have a zero EoS parameter at high redshifts, and then its EoS suddenly transitions to a nonzero constant $w_{\rm EDM,0}$ at some low redshift $z_{t}$. This is a DM model with a discontinuous EoS and two free parameters. By analyzing this model with the DESI BAO, Planck CMB, and SN Ia data, the authors found that it yields minimum chi-squared values very close to that of the $w_0w_a$CDM model. It is worth noting that Ref.~\cite{Braglia:2025gdo} considered only the case where DE is a cosmological constant. To explore the non-cold tendency of DM under different DE scenarios, Ref.~\cite{Li:2025dwz} used BAO data from DESI DR2, CMB data from Planck, and SN Ia data from DESY5 and PantheonPlus (PP) to investigate a constant-EoS-parameter DM model accompanied by the cosmological constant, constant-EoS-parameter DE, and Chevallier-Polarski-Linder (CPL) DE models, respectively. The results showed a preference for a nonzero DM EoS parameter at the $2.8\sigma$ and $3.3\sigma$ levels within the context of the constant-EoS-parameter DE. However, this model was moderately to strongly disfavored by $\Lambda$CDM in model comparison. We attribute this outcome to an inappropriate parameterization of DM, since DM must be very close to CDM in the early universe as constrained by the CMB observations. Apart from addressing the DESI anomaly by introducing NSDM, some researchers treat DE and DM as a unified fluid~\cite{Wang:2024rus,Su:2025ntt,Kou:2025yfr}. This approach can also resolve the DESI anomaly while avoiding violations of the null energy condition, although in this work, we focus exclusively on the NSDM model.

In this work we propose a NSDM model with a continuous EoS and only one free parameter. We aim to analyze how this NSDM deviates from CDM under different DE scenarios using the CMB, DESI BAO, and SN Ia datasets, and to understand the extent to which the introduction of this NSDM helps to avoid the violation of the null energy condition.

This work is organized as follows. In Sec.~\ref{sec:1}, we briefly introduce the models considered in this work. In Sec.~\ref{sec:2}, we present the cosmological data utilized in the analysis. In Sec.~\ref{sec:3}, we report the constraint results and make some relevant discussions. The conclusion is given in Sec.~\ref{sec:4}.

\section{Introduction of the new models}
\label{sec:1}
In this section, we introduce the cosmological models investigated in this work. Before that, we first propose a new parametrization for the EoS of DM. Given the constraints from CMB observations, it is known that in the early universe, DM behaved very similarly to CDM. In other words, the EoS of DM is very close to zero in the neighborhood of point where scale factor $a$ equal to zero. Therefore, we assume that both the EoS of DM $w_{\rm dm}$ and its derivative with respect to the scale factor ${\rm d}w_{\rm dm}/{\rm d}a$ vanish at $a=0$. If we perform a Taylor expansion of this EoS around $a=0$, we have $w_{\rm dm}(a)=\sum_{n=0}^{\infty}w_{\rm n+2}a^{n+2}$.  To ensure sufficient precision in the parameter fitting results, we retain only the first nonzero term of the Taylor series to minimize the total number of free parameters, and thus we obtain $w_{\rm dm}(a)=w_{2}a^{2}$ as the EoS of our NSDM in this work.

Now we consider a spatially flat, homogeneous, and isotropic universe described by the Friedmann-Robertson-Walker (FRW) metric within the framework of general relativity. The matter components are assumed to be minimally coupled to gravity, with no interactions among them other than gravity. Therefore, the dimensionless Hubble parameter is given by
\begin{equation}
	\frac{H^2(a)}{H_0^2} = \Omega_{\rm r0}a^{-4} + \Omega_{\rm dm0}f_{\rm dm}(a) + \Omega_{\rm b0}a^{-3} + \Omega_{\rm de0} f_{\rm de}(a),
\end{equation}
where $a=1/(1+z)$ is the scale factor, $H(a)$ is the Hubble parameter, $\Omega_{\rm r0}$, $\Omega_{\rm dm0}$, $\Omega_{\rm b0}$, and $\Omega_{\rm de0}$ are the present density parameters for radiation, NSDM, baryons, and DE, respectively. $f_{\rm dm}(a)$ represents the normalized $a$-dependent density of NSDM, while $f_{\rm de}(a)$ represents the normalized $a$-dependent density of DE. They are given by
\begin{equation}
	f_{\rm ds}(a) = \exp\left( -3 \int_{1}^{a} \frac{1 + w_{\rm ds}(a')}{a'} {\rm d}a' \right),
\end{equation}
Here, ``ds" denotes the dark sector, referring to either DM or DE. In this work, we consider three representative parameterizations of DE, namely the cosmological constant with $w_{\rm de}=-1$, constant-EoS-parameter DE with $w_{\rm de}=w$, and CPL DE with $w_{\rm de}=w_0+w_a(1-a)$, respectively. We refer to the three models formed by combining the proposed NSDM with these three types of DE as $\Lambda w_2$DM, $ww_2$DM, and $w_0w_aw_2$DM, respectively.

\section{Data sets and methodology}
\label{sec:2}
\begin{table*}[tbp]
	\centering
	\caption{The $1\sigma$ CL fitting results in the $\Lambda$CDM, $\Lambda w_2$DM, $w$CDM, $ww_2$DM, $w_0w_a$CDM, and $w_0w_aw_2$DM models from the CMB+DESI, CMB+DESI+PP, CMB+DESI+Union3, and CMB+DESI+DESY5 data combinations. Here, $H_{0}$ is in units of ${\rm km}~{\rm s}^{-1}~{\rm Mpc}^{-1}$.}
	\label{tab:1}
	\setlength{\tabcolsep}{2mm}
	\renewcommand{\arraystretch}{1.2}
		\footnotesize
		\begin{tabular}{lc c c c c c c}
			\hline 
			\hline
			Model/Dataset & $H_0$ &$\Omega_{\mathrm{m}}$& $w_2$ & $w$ or $w_0$ & $w_a$ \\
			\hline
			$\bm{\Lambda}$\textbf{CDM} &  &  &  &  &  \\
			CMB+DESI & $68.66^{+0.29}_{-0.30}$ & $0.2985^{+0.0038}_{-0.0037}$ & --- & --- & --- \\
			CMB+DESI+PP & $68.55\pm0.29$ & $0.2999\pm0.0037$ & --- & --- & --- \\
			CMB+DESI+Union3 & $68.57\pm0.29$ & $0.2997\pm0.0037$ & --- & --- & --- \\
			CMB+DESI+DESY5 & $68.45\pm0.29$ & $0.3013\pm0.0037$ & --- & --- & --- \\
			\hline
			$\bm{\Lambda w_2\textbf{DM}}$ &  &  &  &  &  \\
			CMB+DESI & $68.80^{+0.31}_{-0.30}$ & $0.2983\pm0.0038$ & $-0.0112\pm0.0048$ & --- & --- \\
		CMB+DESI+PP & $68.66^{+0.30}_{-0.29}$ & $0.3000^{+0.0036}_{-0.0037}$ &$-0.0103\pm0.0048$  & --- & --- \\
		CMB+DESI+Union3 & $68.69^{+0.30}_{-0.31}$ & $0.2998^{+0.0037}_{-0.0038}$ & $-0.0106\pm0.0048$ & --- & --- \\
		CMB+DESI+DESY5 & $68.55^{+0.29}_{-0.30}$ & $0.3014\pm0.0037$ & $-0.0097\pm0.0047$ & --- & --- \\
		\hline
			$\bm{w}$\textbf{CDM} &  &  &  &  &  \\
			CMB+DESI & $69.18^{+0.91}_{-1.00}$ & $0.2948^{+0.0074}_{-0.0075}$ & --- & $-1.022^{+0.040}_{-0.037}$ & --- \\
		CMB+DESI+PP & $68.00^{+0.58}_{-0.57}$ & $0.3036\pm0.005$ & --- & $-0.9738\pm0.0235$ & --- \\
		CMB+DESI+Union3 & $67.89^{+0.69}_{-0.68}$ & $0.3045^{+0.0057}_{-0.0058}$ & --- & $-0.9698\pm0.0283$ & --- \\
		CMB+DESI+DESY5 & $67.37^{+0.55}_{-0.54}$ & $0.3085\pm0.0049$ & --- & $-0.9485^{+0.0218}_{-0.0221}$ & --- \\
		\hline
			$\bm{{ww_2}\textbf{DM}}$ &  &  &  &  &  \\
			CMB+DESI & $67.79^{+1.00}_{-1.12}$ & $0.3061^{+0.0085}_{-0.0087}$ & $-0.0148^{+0.0062}_{-0.0061}$ & $-0.9549^{+0.0463}_{-0.00464}$ & --- \\
		CMB+DESI+PP & $67.42^{+0.63}_{-0.62}$ & $0.3088^{+0.0055}_{-0.0056}$ & $-0.0159\pm0.0055$ & $-0.9387^{+0.0273}_{-0.0274}$ & --- \\
		CMB+DESI+Union3 & $66.89\pm0.76$ & $0.3130^{+0.0066}_{-0.0065}$ & $-0.0179^{+0.0057}_{-0.0058}$ & $-0.9159^{+0.0332}_{-0.0334}$ & --- \\
		CMB+DESI+DESY5 & $66.72^{+0.60}_{-0.59}$ & $0.3144\pm0.0054$ & $-0.0184^{+0.0055}_{-0.0054}$ & $-0.9088^{+0.0261}_{-0.0259}$ & --- \\
		\hline
			$\bm{w_0w_a}$\textbf{CDM} &  &  &  &  &  \\
			CMB+DESI & $63.58\pm2.08$ & $0.3545^{+0.0228}_{-0.0257}$ & --- & $-0.3882^{+0.2281}_{-0.2599}$ & $-1.816^{+0.789}_{-0.646}$ \\
		CMB+DESI+PP & $67.70\pm0.59$ & $0.3096^{+0.0057}_{-0.0056}$ & --- & $-0.8456^{+0.0543}_{-0.546}$ & $-0.5398^{+0.2271}_{-0.2031}$ \\
		CMB+DESI+Union3  & $66.12^{+0.84}_{-0.83}$ & $0.3256\pm0.0087$ & --- & $-0.6752^{+0.0886}_{-0.889}$ & $-1.0179^{+0.3229}_{-0.2843}$ \\
		CMB+DESI+DESY5 & $66.94\pm0.57$ & $0.3173^{+0.0058}_{-0.0057}$ & --- & $-0.7589^{+0.0582}_{-0.578}$ & $-0.7946^{+0.2477}_{-0.2231}$ \\
		\hline
			$\bm{{w_0w_aw_2}\textbf{DM}}$ &  &  &  &  &  \\
			CMB+DESI & $64.28\pm2.34$ &  $0.3460^{+0.0242}_{-0.0296}$& $-0.0048^{+0.0082}_{-0.0065}$ & $-0.4969^{+0.2630}_{-0.3185}$ & $-1.438^{+1.017}_{-0.794}$ \\
		CMB+DESI+PP & $67.45^{+0.61}_{-0.62}$ & $0.3104\pm0.0057$ & $-0.0115^{+0.0074}_{-0.0063}$ & $-0.8856^{+0.0591}_{-0.0589}$ & $-0.2663^{+0.2594}_{-0.2493}$ \\
		CMB+DESI+Union3 & $66.14^{+0.83}_{-0.84}$ & $0.3245\pm0.0088$ & $-0.0075^{+0.0071}_{-0.0061}$ & $-0.7235^{+0.0971}_{-0.0973}$ & $-0.7693^{+0.3782}_{-0.3456}$ \\
		CMB+DESI+DESY5 & $66.79\pm0.58$ & $0.3176\pm0.0057$ & $-0.0087^{+0.0070}_{-0.0061}$ & $-0.7963^{+0.0622}_{-0.0623}$ & $-0.5573^{+0.2833}_{-0.2819}$ \\
		\hline
			\hline
		\end{tabular}
	\end{table*}

In this section, we list the observational datasets used to constrained the free parameters of the three proposed models, and introduce the methods and tools used for parameter extraction and data analysis.

For the CMB data, since the NSDM adopted in this work closely approximates CDM at high redshift, the early-time evolution shows almost no deviation from $\Lambda$CDM. Therefore, we can adopt the Planck2018 CMB distance priors provided in Table F1 of Ref.~\cite{Rubin:2023jdq} instead of using the full temperature anisotropy and polarization power spectrum dataset. The distance priors is consist of three parameters, the shift parameter $R$, the acoustic angular scale $\theta_{\ast}$, and the baryon density $\omega_{\rm b}=\Omega_{\rm b0}h^2$, where $h=\frac{H_0}{\rm 100km/s/Mpc}$.

The shift parameter $R$ reads
\begin{equation}
	R=\sqrt{\Omega_{\rm m0}H_0^2}D_{\rm M}(z_{\ast}),
\end{equation}
here $\Omega_{\rm m0}=\Omega_{\rm dm0}+\Omega_{\rm b0}$, and $D_{\rm M}(z_{\ast})$ is the comoving angular diameter distance at the recombination redshift $z_{\ast}$, which depends on the dominant components after decoupling and is defined by
\begin{equation}\label{}
	D_{\rm M}(z_{\ast})=(1+z_{\ast})D_{\rm A}(z_{\ast})=\int_{0}^{z_{\ast}}\frac{{\rm d}z}{H(z)},
\end{equation}
the acoustic angular scale $\theta_{\ast}$ reads
\begin{equation}
	 \theta_{\ast}=\frac{r_{\rm s}(z_{\ast})}{D_{\rm M}(z_{\ast})},
\end{equation}
in which $r_{\rm s}(z_{\ast})$ is the sound horizon at the recombination redshift $z_{\ast}$, which depends on the dominant components before decoupling and have the following form
\begin{equation}\label{}
	r_{\rm s}(z_{\ast})=\int_{z_{\ast}}^{\infty}\frac{c_{\rm s}(z){\rm d}z}{H(z)},
\end{equation}
the sound speed $c_{\rm s}(z)=1/\sqrt{3(1+\bar{R}_{\rm b}/(1+z))}$, $\bar{R}_{\rm b }=3\omega_{\rm b}/4\times2.469\times10^{-5}$.
The recombination redshift $z_{\ast}$ is fitted as \cite{hu1996small}
\begin{eqnarray}
	z_{*}&=&1048(1+0.00124\omega_{\rm b}^{-0.738})(1+g_{1}\omega_{\rm m}^{g_{2}}),\\
	g_{1}&=&\frac{0.0783\omega_{\rm b}^{-0.238}}{1+39.5\omega_{\rm b}^{0.763}},\\
	g_{2}&=&\frac{0.56}{1+21.1\omega_{\rm b}^{1.81}},
\end{eqnarray}
here $\omega_{\rm m}=\Omega_{\rm m0}h^2$.

For the BAO data, we use the BAO measurements from the Data Release (DR) 2 of DESI. It provides the most precise BAO constraints to date, covering a wide redshift range $(0.1 < z < 4.2)$ with different galaxy and quasar samples. DESI BAO measures the volume-averaged distance $D_{\rm V}(z)=[D^2_{\rm M}(z)z/H(z)]^{\frac{1}{3}}$, comoving angular diameter distance $D_{\rm M}(z)$, and Hubble distance $D_{\rm H}(z)=1/H(z)$ in terms of $D_{\rm V}(z)/r_{\rm d}$, $D_{\rm M}(z)/r_{\rm d}$, and $D_{\rm H}(z)/r_{\rm d}$ at seven different redshift. $r_{\rm d}=r_{\rm s}(z_{\rm d})$ is the sound horizon at the drag epoch $z_{\rm d}$, the drag epoch redshift $z_{\rm d}$ is fitted as \cite{hu1996small}
\begin{eqnarray}
	z_{\rm d} &=&1291\frac{\omega_{\rm m}^{0.251}}{1+0.659\omega_{\rm m}^{0.828}}\frac{1+b_{1}\omega_{\rm b}^{b_{2}}}{0.962} \\
	b_1 &=& 0.313\omega_{\rm m}^{-0.419}(1+0.607\omega_{\rm m}^{0.674})\\
	b_2&=&0.238\omega_{\rm m}^{0.223}
\end{eqnarray}

For SN Ia data, we employ three independent SN Ia samples in our analysis: the Dark Energy Survey Year 5 (DESY5) compilation includes 1,635 well-measured SN Ia in the redshift range $0.10 < z < 1.13$, together with an external low-redshift sample of 194 SNe Ia spanning $0.025 < z < 0.10$~\cite{DES:2024jxu}, the Union3 compilation which consists of 2,087 SN Ia spanning the redshift range $0.05 < z < 2.26$~\cite{Rubin:2023jdq}, and the PP sample consists of 1,550 SN Ia spanning the redshift range $0.01 < z < 2.26$~\cite{scolnic2022pantheon+}. 

We perform Markov Chain Monte Carlo (MCMC) simulations using the publicly available Python code emcee~\cite{foreman2013emcee} and analyze the resulting samples with the GetDist Python package~\cite{Lewis:2019xzd}. We consider the following combinations of datasets: CMB+DESI, CMB+DESI+PP, CMB+DESI+Union3, and CMB+DESI+DESY5, in order to derive the mean values, confidence intervals, and posterior distributions of the model parameters for the three proposed models. Moreover, we assign a flat prior to each model parameter. Since these three proposed models are extensions of $\Lambda$CDM, we can derive their significance relative to $\Lambda$CDM by computing the difference between their minimum chi-squared values and that of $\Lambda$CDM.

\section{Results and discussion}
\label{sec:3}
\begin{figure}[!htbp]
	\centering
	\includegraphics[scale=0.45]{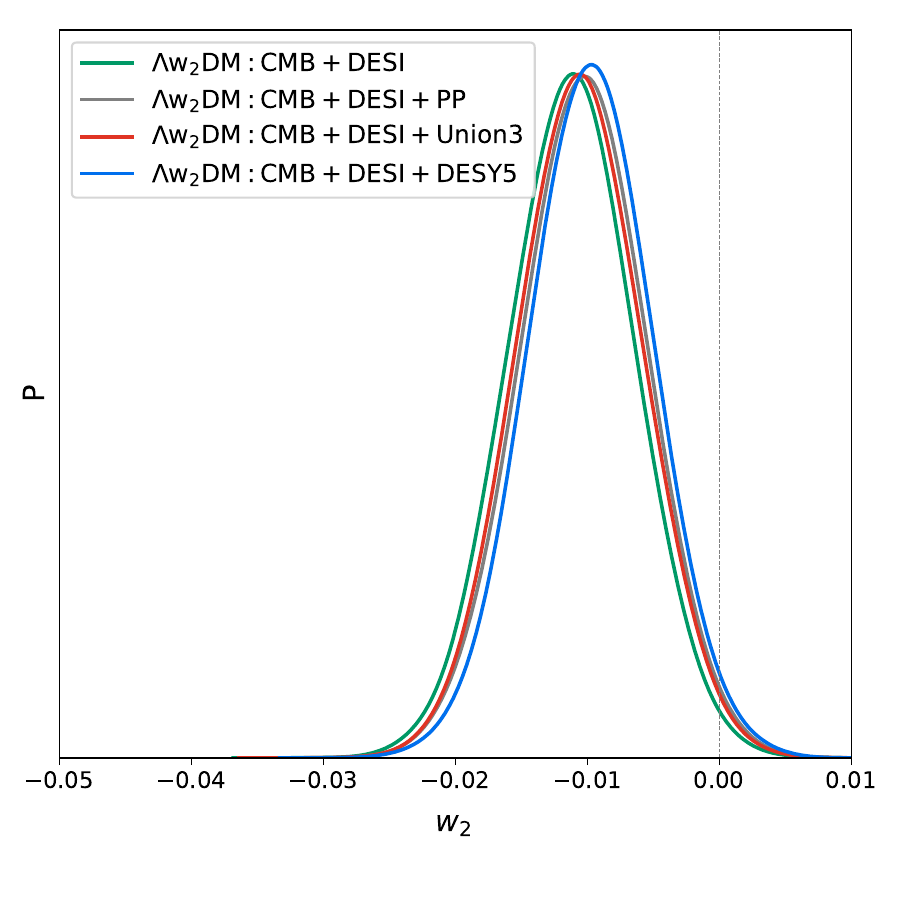}
	\caption{The one-dimensional marginalized posterior distributions of $w_2$ for the $\Lambda w_2$DM model using CMB+DESI, CMB+DESI+PP, CMB+DESI+Union3, and CMB+DESI+DESY5 data combinations. The dashed line corresponds to $w_2=0$. }
	\label{fig:1}
\end{figure}

\begin{figure*}[!htbp]
	\begin{minipage}{0.45\linewidth}
		\centerline{\includegraphics[width=1\textwidth]{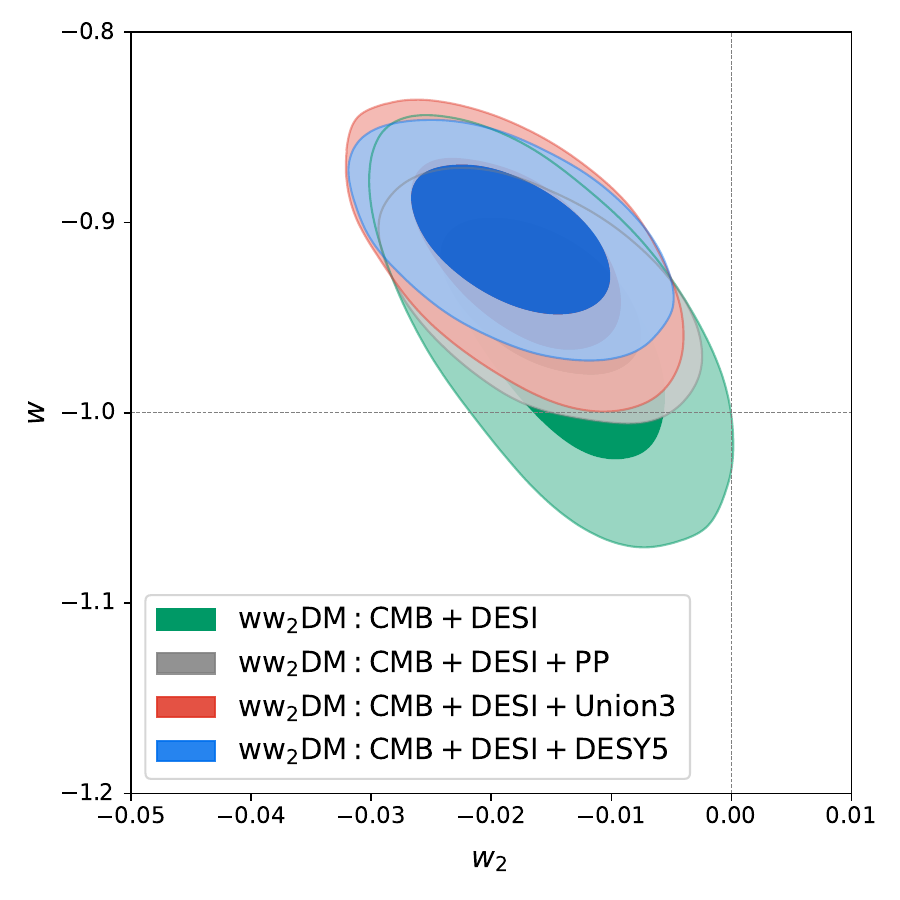}}
	\end{minipage}
	\begin{minipage}{0.45\linewidth}
		\centerline{\includegraphics[width=1\textwidth]{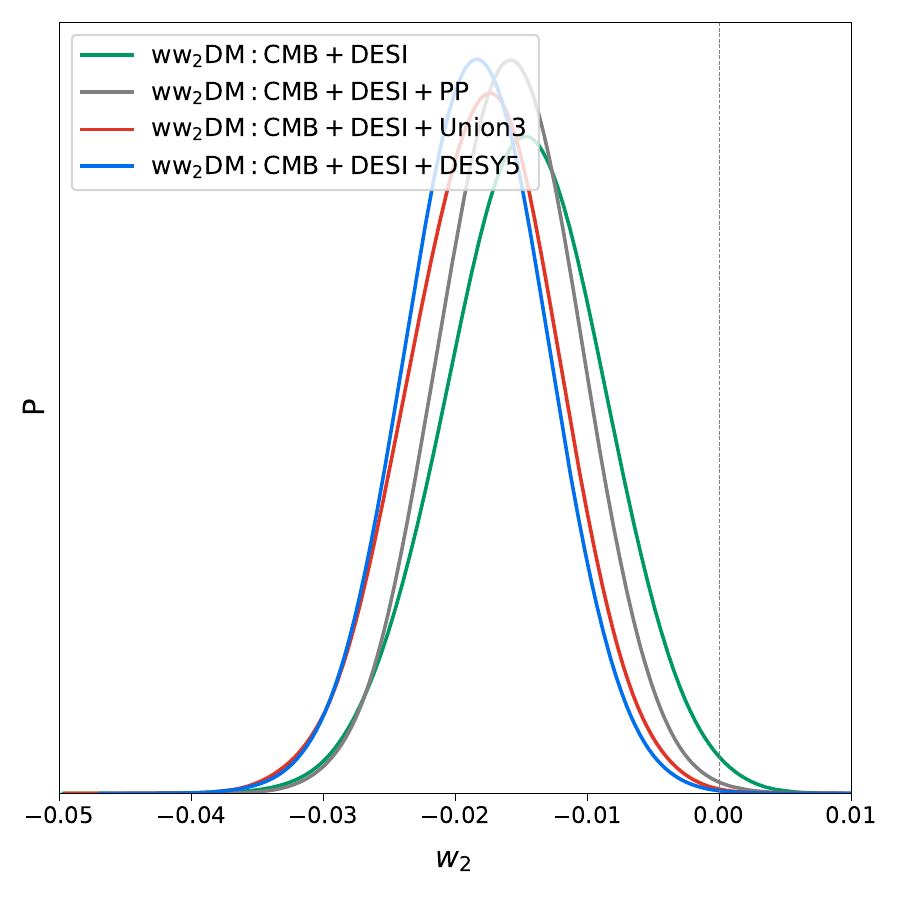}}
	\end{minipage}
	\caption{The two-dimensional joint contours at $1\sigma$ and $2\sigma$ CL of parameters $w$ and $w_2$ (left panel) and one-dimensional marginalized posterior distributions of $w_2$ (right panel) for the $ww_2$DM model using CMB+DESI, CMB+DESI+PP, CMB+DESI+Union3, and CMB+DESI+DESY5 data combinations. The horizontal dashed line corresponds to $w=-1$, and two vertical dashed lines corresponds to $w_2=0$.}
	\label{fig:2}
\end{figure*}
\begin{figure*}[btp]
	\centering
	\includegraphics[scale=0.7]{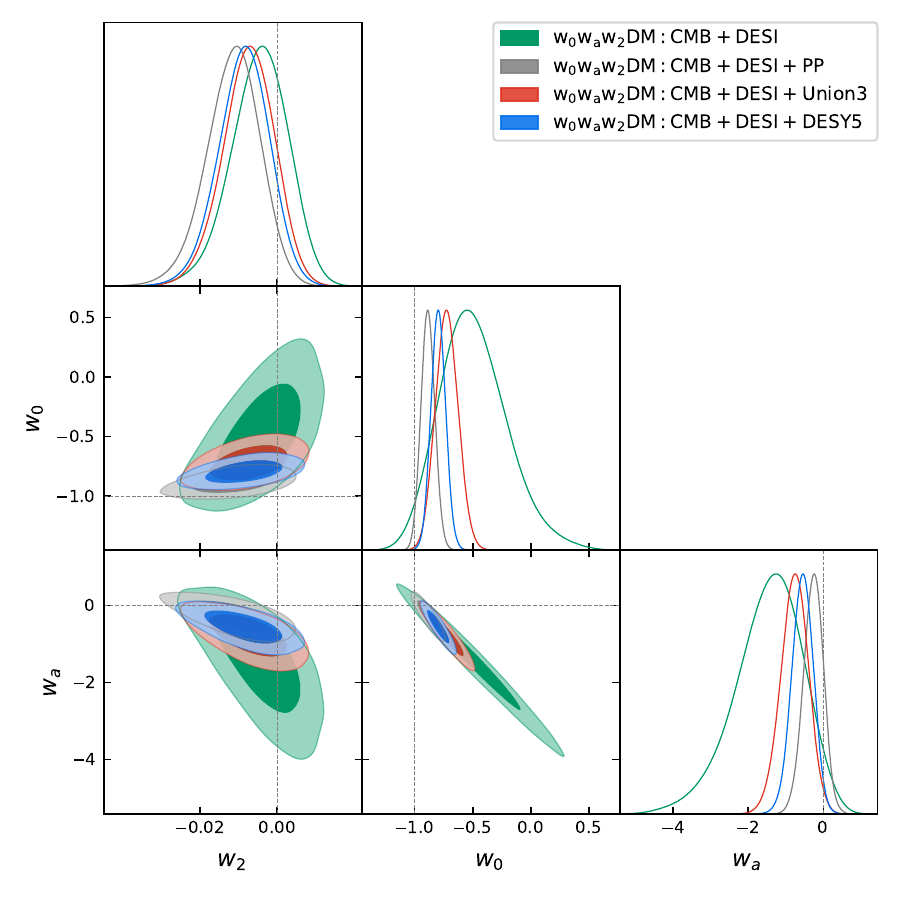}
	\caption{The triangular plot of the fitting results for the $w_0w_aw_2$DM model using CMB+DESI, CMB+DESI+PP, CMB+DESI+Union3, and CMB+DESI+DESY5 data combinations. The dashed lines correspond to the three cases: $w_0=-1$, $w_a=0$, and $w_2=0$. }
	\label{fig:3}
\end{figure*}
\begin{figure*}
	\begin{minipage}{0.45\linewidth}
		\centerline{\includegraphics[width=1\textwidth]{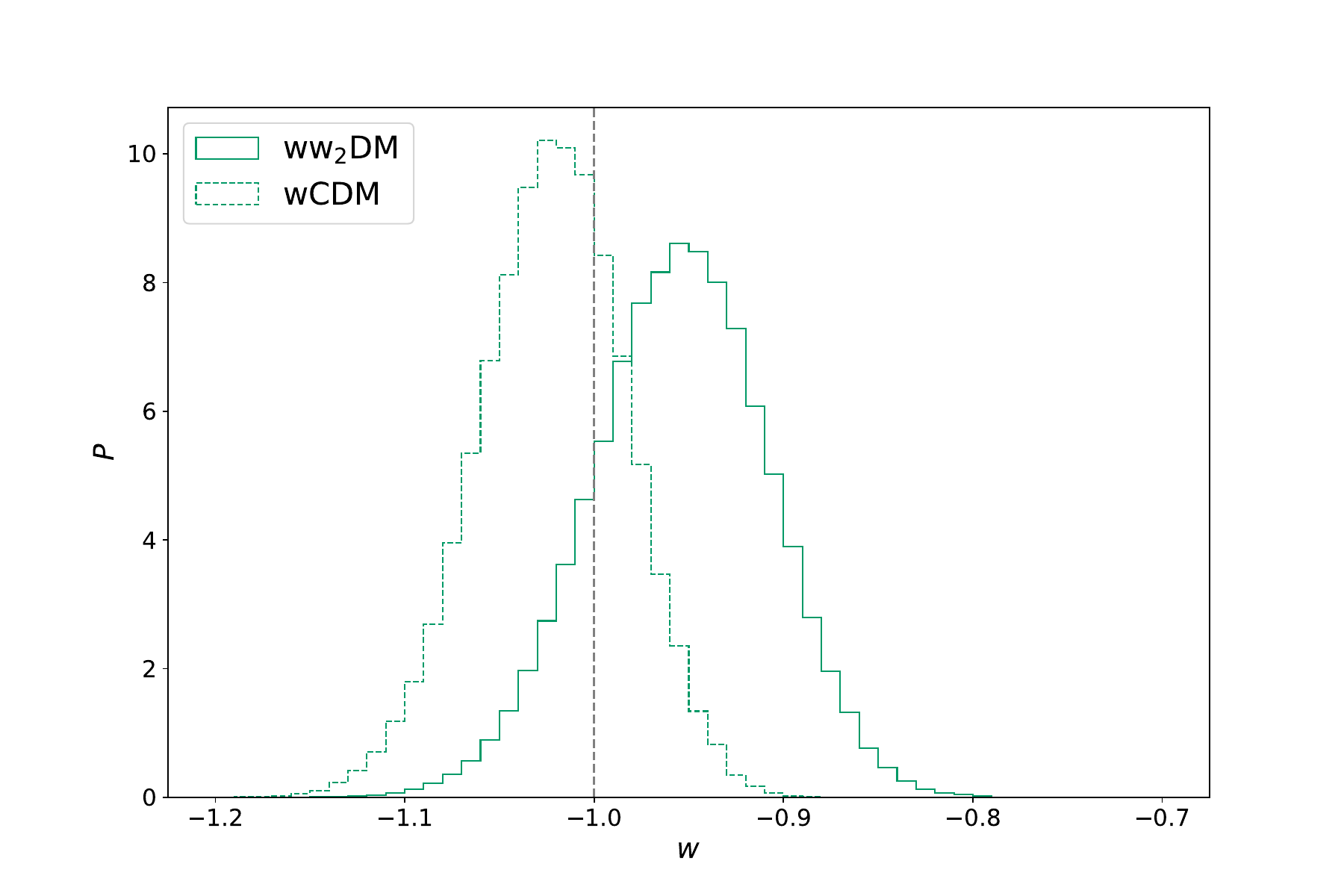}}
	\end{minipage}
	\begin{minipage}{0.45\linewidth}
		\centerline{\includegraphics[width=1\textwidth]{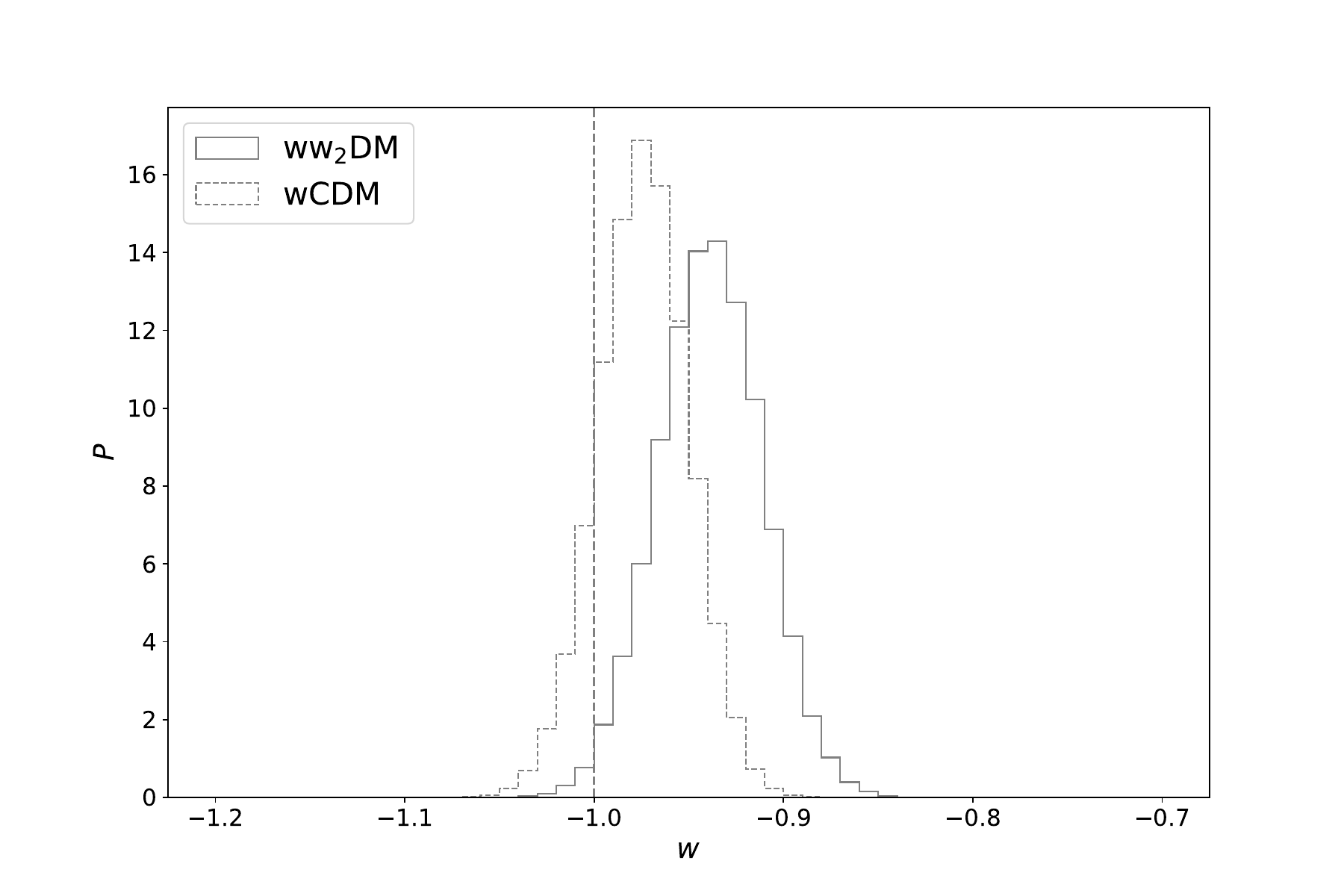}}
	\end{minipage}
	\begin{minipage}{0.45\linewidth}
		\centerline{\includegraphics[width=1\textwidth]{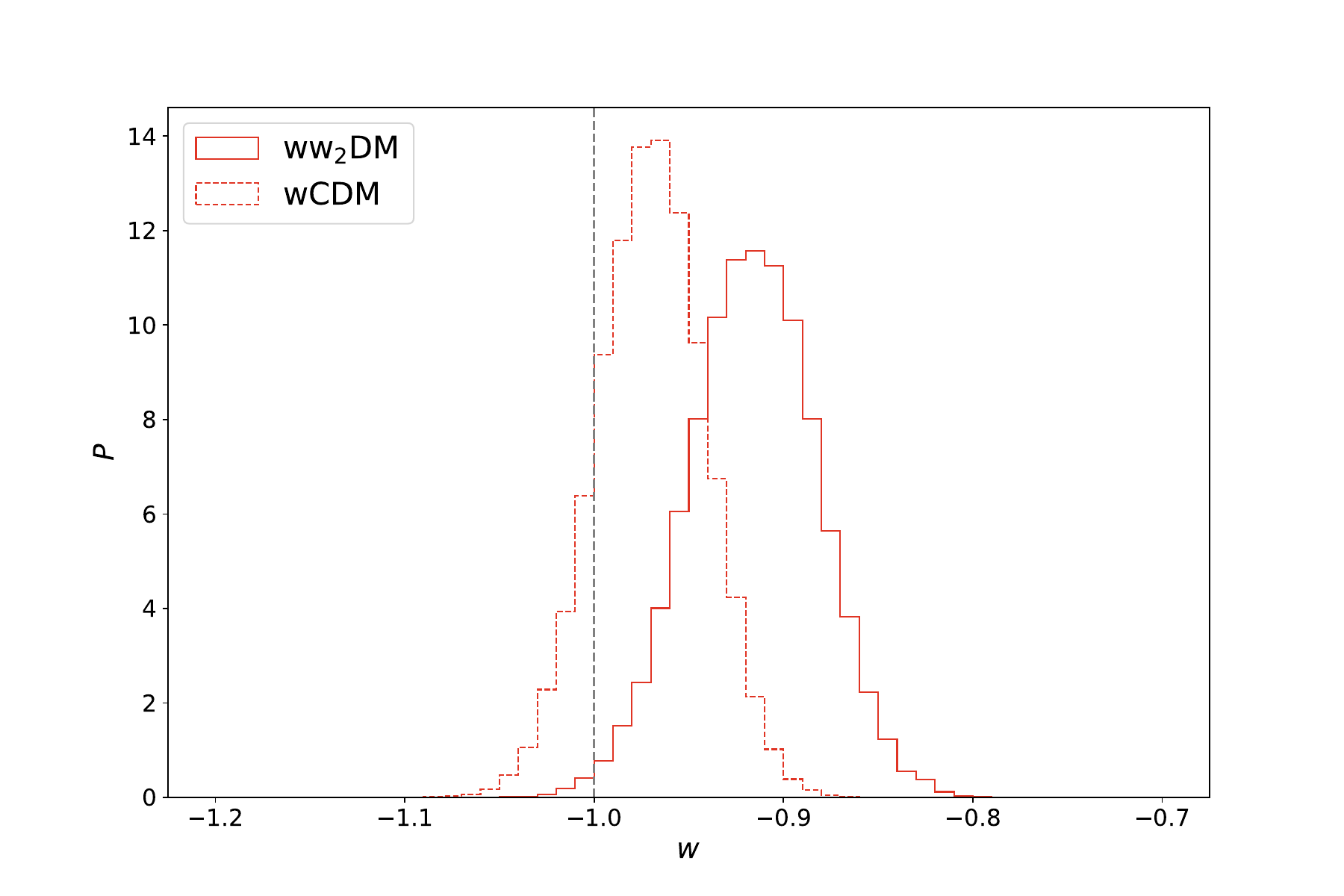}}
	\end{minipage}
	\begin{minipage}{0.45\linewidth}
		\centerline{\includegraphics[width=1\textwidth]{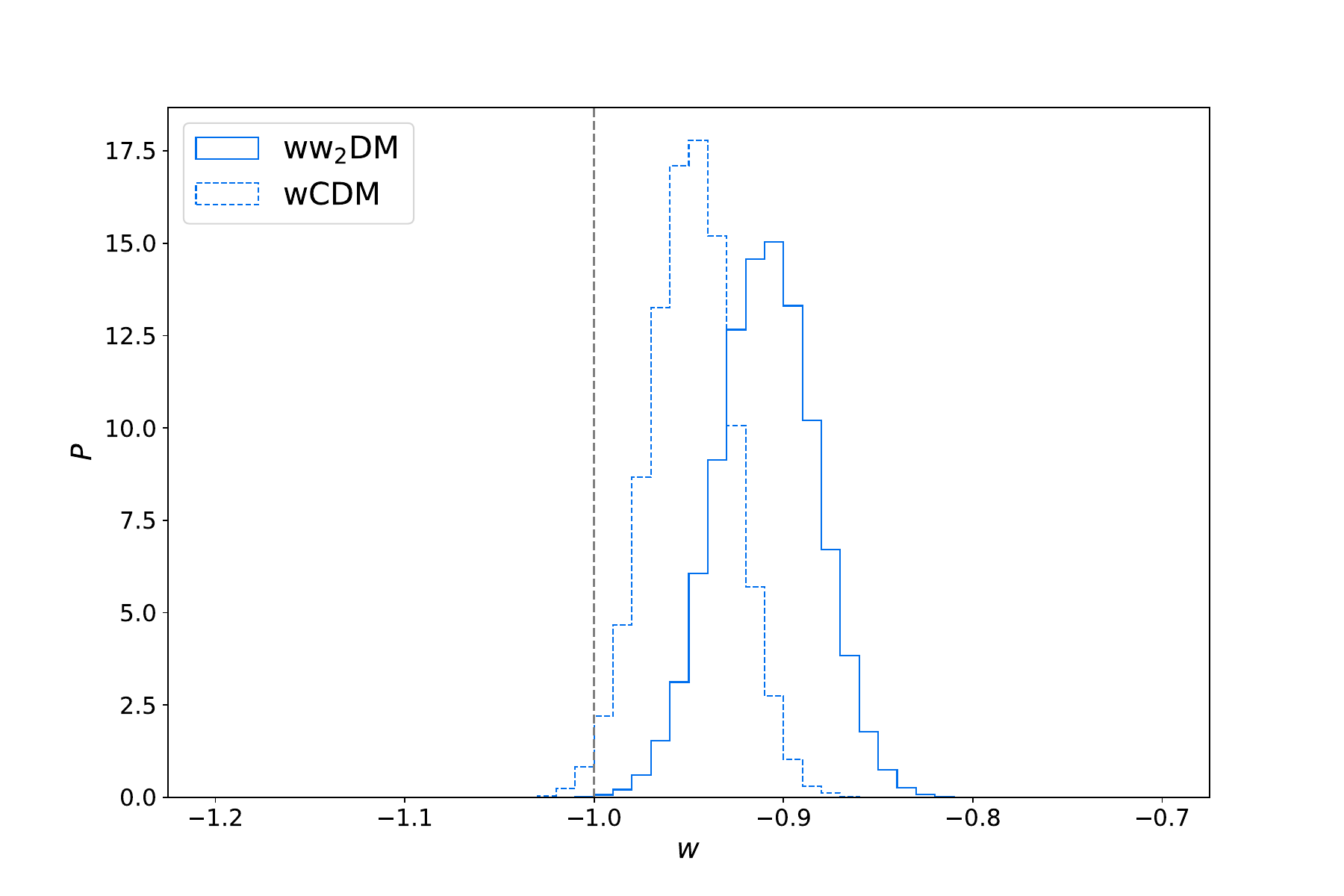}}
	\end{minipage}
	\caption{The binned one-dimensional marginalized posterior distributions of $w$ for the $w$CDM and $ww_2$DM models under CMB+DESI (top left), CMB+DESI+PP (top right), CMB+DESI+Union3 (bottom left), and CMB+DESI+DESY5 (bottom right) data combinations. The dashed line corresponds to $w=0$.}
	\label{fig:4}
\end{figure*}
\begin{figure*}
	\begin{minipage}{0.45\linewidth}
		\centerline{\includegraphics[width=1\textwidth]{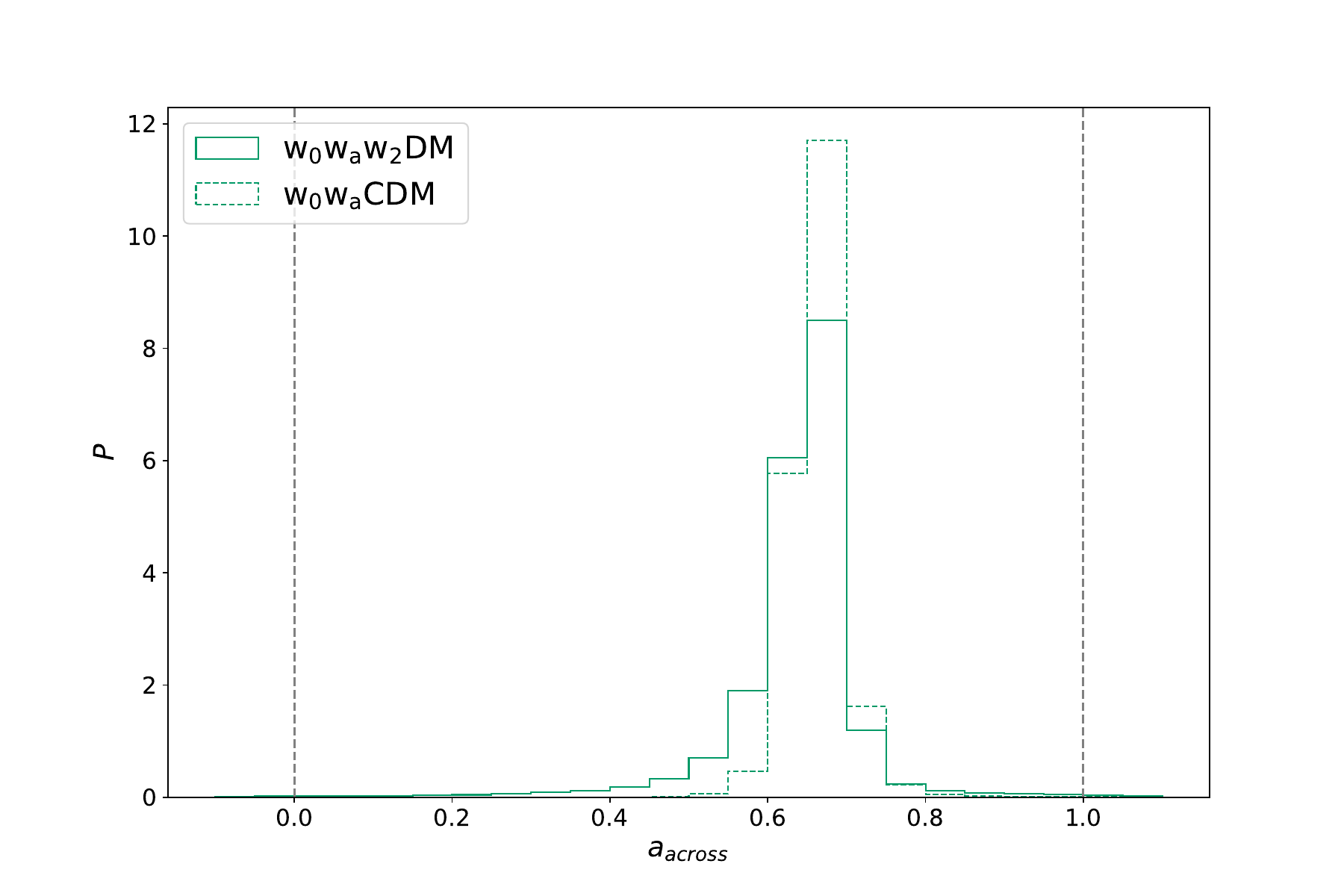}}
	\end{minipage}
	\begin{minipage}{0.45\linewidth}
		\centerline{\includegraphics[width=1\textwidth]{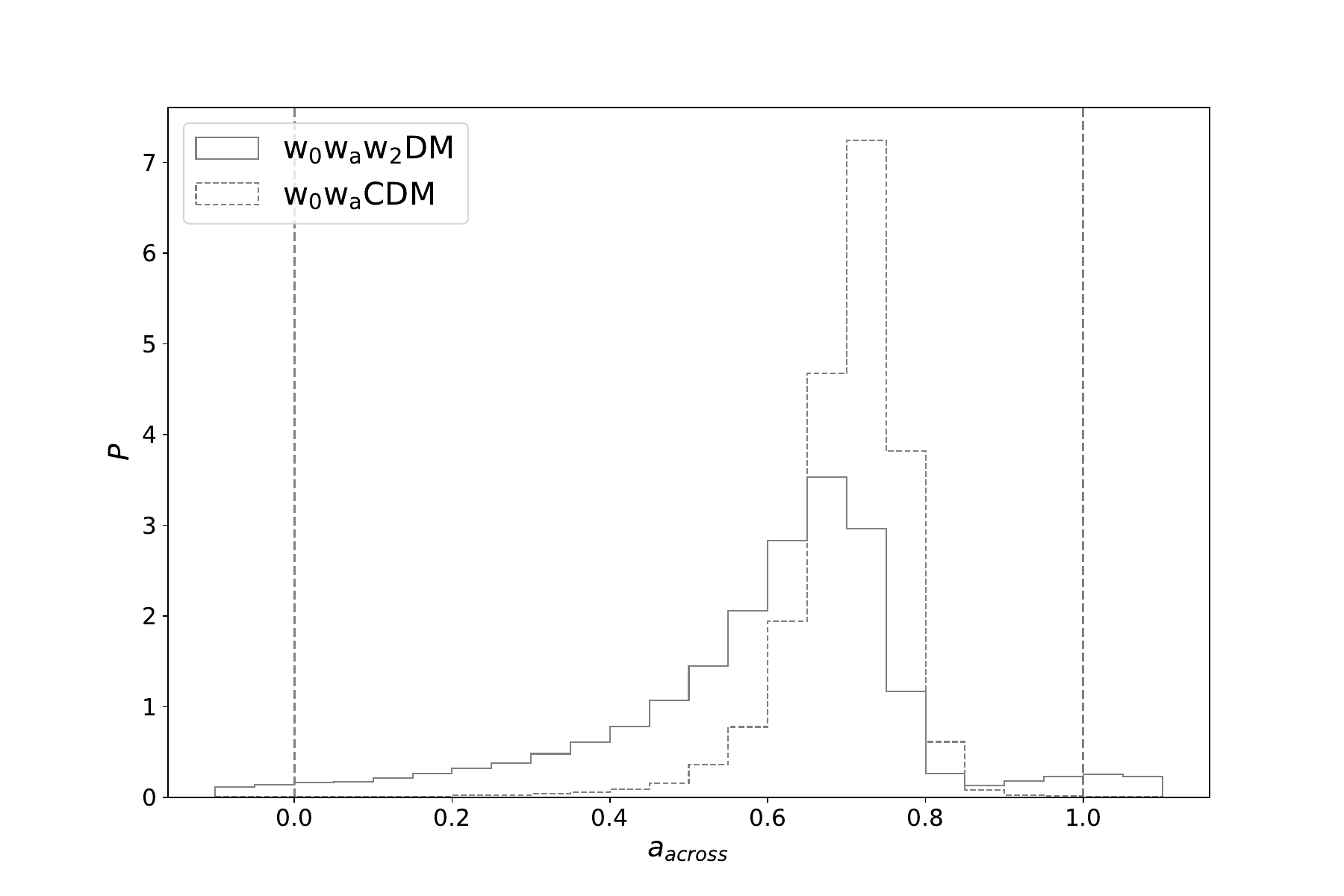}}
	\end{minipage}
	\begin{minipage}{0.45\linewidth}
		\centerline{\includegraphics[width=1\textwidth]{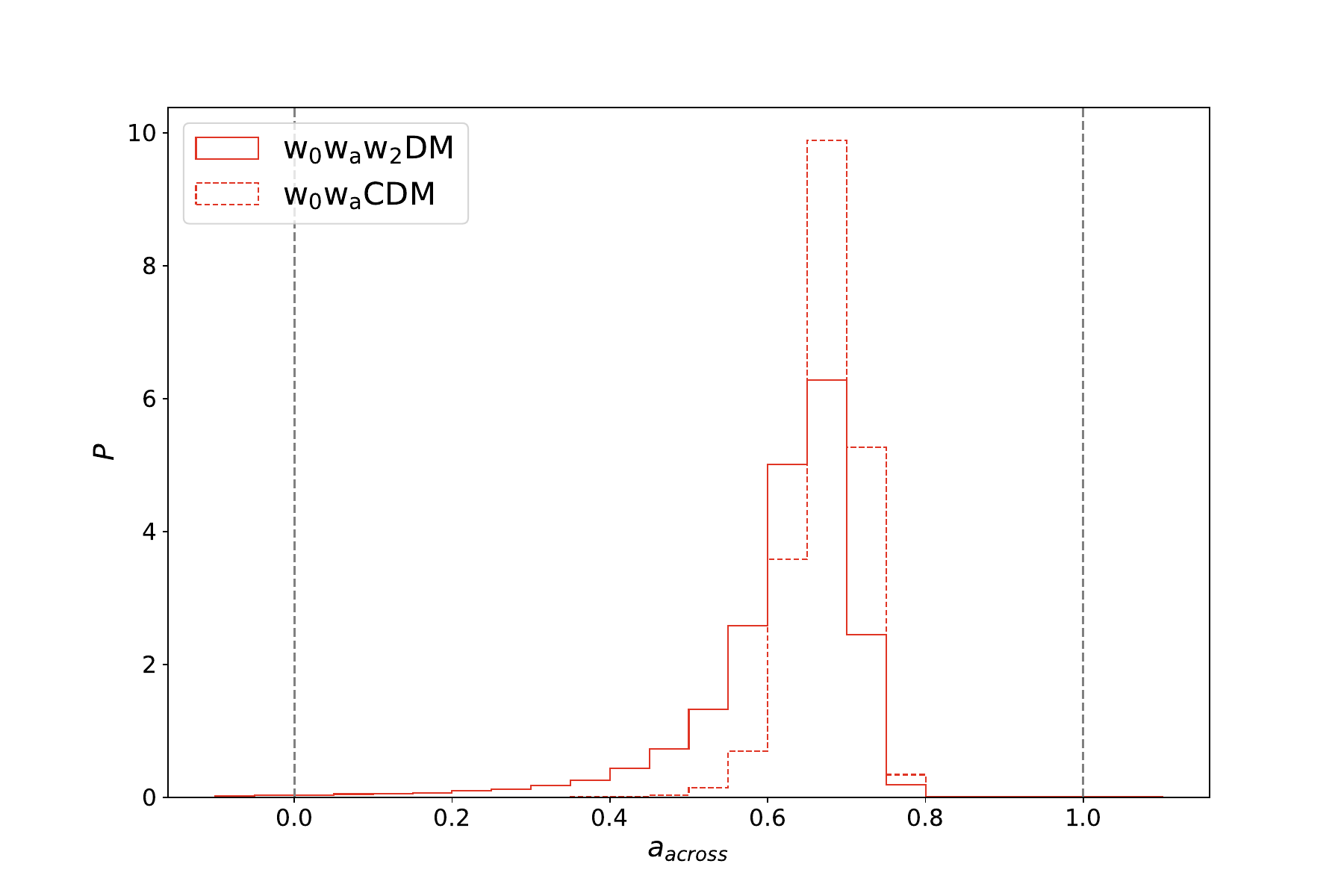}}
	\end{minipage}
	\begin{minipage}{0.45\linewidth}
		\centerline{\includegraphics[width=1\textwidth]{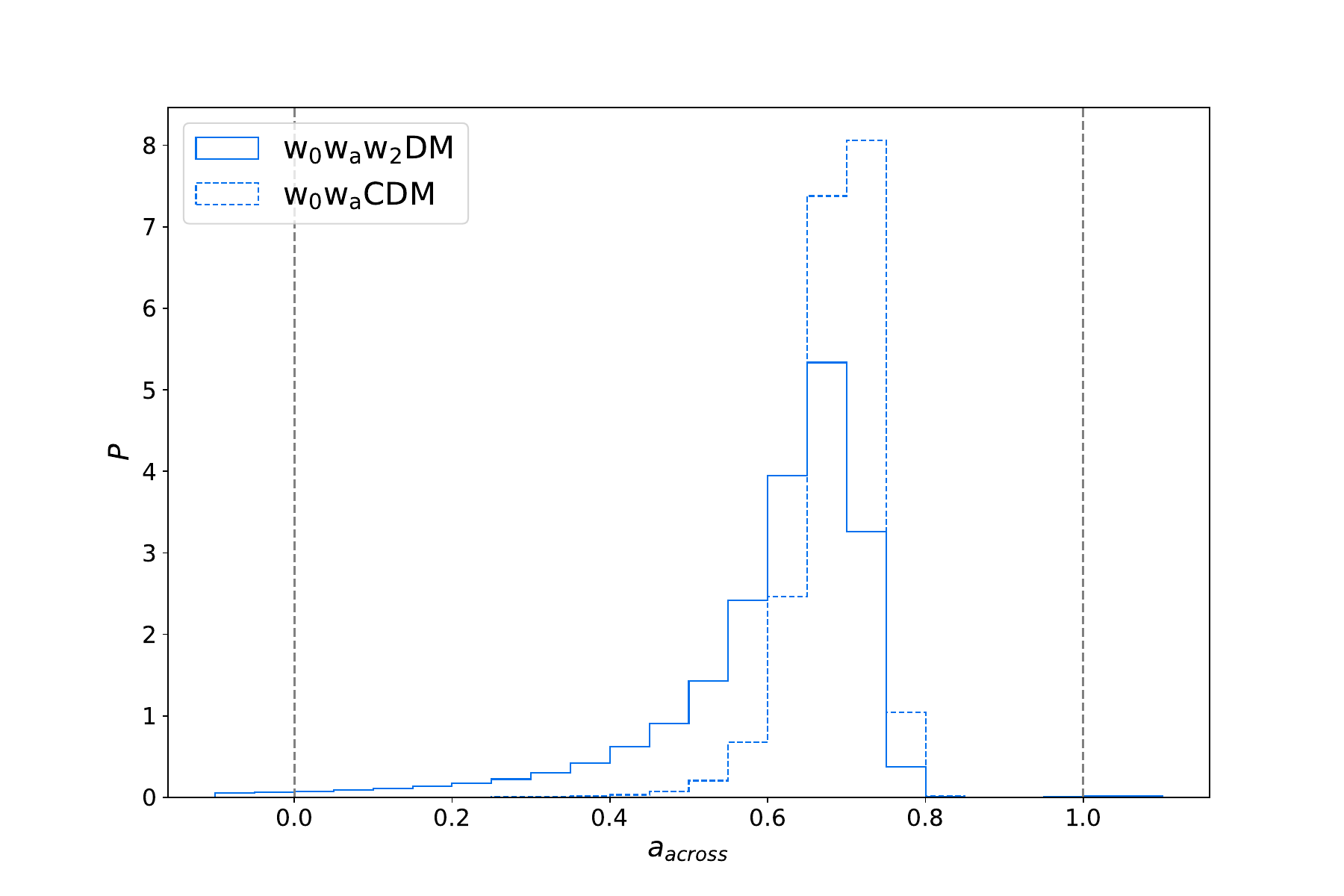}}
	\end{minipage}
	\caption{The binned one-dimensional marginalized posterior distributions of the crossing scale factor $a_{\mathrm{across}}$ for the $w_0w_a$CDM and $w_0w_aw_2$DM models under CMB+DESI (top left), CMB+DESI+PP (top right), CMB+DESI+Union3 (bottom left), and CMB+DESI+DESY5 (bottom right) data combinations. two dashed lines corresponds to $a_{\mathrm{across}}=0$ and $a_{\mathrm{across}}=1$, respectively.}
	\label{fig:5}
\end{figure*}

In this section, we present the constraints on the cosmological parameters. We consider the $\Lambda$CDM, $\Lambda w_2$CDM, $w$CDM, $ww_2$DM, $w_0w_a$CDM, and $w_0w_aw_2$DM models to perform a cosmological analysis using current observational datasets, including the Planck2018 CMB distance priors, DESI DR2 BAO, PP, Union3, and DESY5 data. We show the one-dimensional marginalized posterior distributions and two-dimensional joint contours at $1\sigma$ and $2\sigma$ confidence levels (CL) for the most relevant parameters in the $\Lambda w_2$DM, $ww_2$DM, and $w_0w_aw_2$DM models in Figs.~\ref{fig:1}-\ref{fig:3}. The marginalized parameter constraints with $1\sigma$ errors are summarized in Tab.~\ref{tab:1}. Furthermore, we compare the binned one-dimensional marginalized posterior distributions of the DE EoS parameter $w$ between the $w$CDM and $ww_2$DM models, as shown in Fig.~\ref{fig:4}, and the across scale factor $a_{\rm across}$ between the $w_0w_a$CDM and $w_0w_aw_2$DM models, as shown in Fig.~\ref{fig:5}, to evaluate the extent to which introducing this NSDM can alleviate the violation of the null energy condition. Finally, we list in Tab.~\ref{tab:2} the minimum chi-square differences between the other models and the $\Lambda$CDM model, along with the corresponding statistical significance levels indicating how much better these models perform compared to $\Lambda$CDM.

In Fig.~\ref{fig:1}, we present the one-dimensional marginalized posterior distributions of $w_2$ for the $\Lambda w_2$DM model using different combinations of datasets. The fitting results of $w_2$ are $-0.0112\pm0.0048$ (CMB+DESI), $-0.0103\pm0.0048$ (CMB+DESI+PP), $-0.0106\pm0.0048$ (CMB+DESI+Union3), and $-0.0097\pm0.0047$ (CMB+DESI+DESY5), indicating a preference for a non-zero DM EoS parameter at $2.3\sigma$, $2.3\sigma$, $2.2\sigma$, and $2.0\sigma$ CL, respectively. We can see that in the $\Lambda w_2$DM model, the NSDM exhibits a negative EoS, which implies that DM  possesses exotic properties similar to those of DE at low redshifts. However, since its EoS remains larger than $-1/3$, it cannot drive the accelerated expansion of the universe by itself.

In Fig.~\ref{fig:2}, we present the two-dimensional joint contours at $1\sigma$ and $2\sigma$ CL of parameters $w$ and $w_2$ on the left panel and one-dimensional marginalized posterior distributions of $w_2$ on the right panel for the $ww_2$DM model using different combinations of datasets. The fitting results of $w_2$ are $-0.0148^{+0.0062}_{-0.0061}$ (CMB+DESI), $-0.0159\pm0.0055$ (CMB+DESI+PP), $-0.0179^{+0.0057}_{-0.0058}$ (CMB+DESI+Union3), and $-0.0184^{+0.0055}_{-0.0054}$ (CMB+DESI+DESY5), indicating a preference for a non-zero DM EoS parameter at $2.4\sigma$, $2.9\sigma$, $3.1\sigma$, and $3.3\sigma$ CL, respectively. We can see that in the $\ ww_2$DM model, the NSDM still exhibits a negative EoS, moreover, the deviation from zero becomes even larger. This is because there exists a negative correlation between $w$ and $w_2$ in this case, and all datasets yield a mean value of $w>-1$, as shown in the left panel of Fig.~\ref{fig:2}.

In Fig.~\ref{fig:3}, we show the triangular plot of the fitting results for the $w_0w_aw_2$DM model using different combinations of datasets. We find that for CMB+DESI, $w_2 = -0.0048^{+0.0082}_{-0.0065}$, showing no evidence for a non-zero DM EoS parameter. When the SN Ia data are included, we obtain $w_2 = 0.0115^{+0.0074}_{-0.0063}$ (CMB+DESI+PP), $w_2 = -0.0075^{+0.0071}_{-0.0061}$ (CMB+DESI+Union3), and $w_2 = -0.0087^{+0.0070}_{-0.0061}$ (CMB+DESI+DESY5), indicating a preference for a non-zero DM EoS parameter at the $1.6\sigma$, $1.1\sigma$, and $1.3\sigma$ CL, respectively. It can be seen that considering the CPL DE sector weakens the non-cold behavior of DM  compared to the cases with the other two DE models. This result is very similar to that reported in Ref.~\cite{Li:2025dwz}.

In Fig.~\ref{fig:4}, we present the binned one-dimensional marginalized posterior distributions of $w$ for the $w$CDM and $ww_2$DM models under the considered data combinations.  Violating the null energy condition requires $w < -1$. Therefore, the probability of violating the null energy condition can be estimated as the ratio of the area under the $w$ distribution where $w < -1$, denoted as $A_{w<-1}$, to the total area under the $w$ distribution, denoted as $A_{\rm tot}$. Accordingly, the significance levels for the $w$CDM model violating the null energy condition are $0.6\sigma$ (CMB+DESI), $-1.1\sigma$ (CMB+DESI+PP), $-1.1\sigma$ (CMB+DESI+Union3), and $-2.3\sigma$ (CMB+DESI+DESY5), respectively. The corresponding significance levels for the $ww_2$DM model are $-1.0\sigma$ (CMB+DESI), $-2.2\sigma$ (CMB+DESI+PP), $-2.5\sigma$ (CMB+DESI+Union3), and $-3.5\sigma$ (CMB+DESI+DESY5), respectively. The formula used to calculate the significance $N_{\sigma}$ is given by $\frac{A_{w<-1}}{A_{\rm tot}}=\int_{-\infty}^{N_{\sigma}}\frac{1}{\sqrt{2\pi}}e^{-\frac{x^2}{2}}{\rm d}x.$  $N_{\sigma}<0$ means that the probability of violating the null energy condition is smaller than 50\%. We can see that, for any combination of datasets, introducing NSDM significantly reduces the probability of violating the null energy condition.

In Fig.~\ref{fig:5}, we present the binned one-dimensional marginalized posterior distributions of the crossing scale factor $a_{\mathrm{across}}$, which is determined by the condition $w_0 + w_a(1 - a_{\mathrm{across}}) = -1$, for the $w_0w_a$CDM and $w_0w_aw_2$DM models under the considered data combinations. Violating the null energy condition requires $0<a_{\mathrm{across}} < 1$. Therefore, the probability of violating the null energy condition can be estimated as the ratio of the area under the $a_{\mathrm{across}}$ distribution where $0<a_{\mathrm{across}} < 1$, denoted as $A_{0<a_{\mathrm{across}}<1} $ to the total area under the $a_{\mathrm{across}}$ distribution, denoted as $A_{\rm tot}$. Accordingly, the significance levels for the $w_0w_a$CDM model violating the null energy condition are $3.0\sigma$ (CMB+DESI), $2.7\sigma$ (CMB+DESI+PP), $4.5\sigma$ (CMB+DESI+Union3), and $3.8\sigma$ (CMB+DESI+DESY5), respectively. The corresponding significance levels for the $w_0w_aw_2$DM model are $2.2\sigma$ (CMB+DESI), $1.2\sigma$ (CMB+DESI+PP), $2.1\sigma$ (CMB+DESI+Union3), and $1.9\sigma$ (CMB+DESI+DESY5), respectively. The formula used to calculate the significance $N_{\sigma}$ is given by $\frac{A_{0<a_{\mathrm{across}}<1}}{A_{\rm tot}}=\int_{-\infty}^{N_{\sigma}}\frac{1}{\sqrt{2\pi}}e^{-\frac{x^2}{2}}{\rm d}x.$  We can see that, in this case, for any combination of datasets, introducing NSDM also significantly reduces the probability of violating the null energy condition.

\begin{table}[!htbp]
	\centering
	\caption{The minimum chi-square differences $\Delta \chi^2_{\rm min} $ of the $\Lambda$CDM model  relative to the $\Lambda w_2$DM, $w$CDM, $ww_2$DM, $w_0w_a$CDM, and $w_0w_aw_2$DM models under CMB+DESI, CMB+DESI+PP, CMB+DESI+Union3, and CMB+DESI+DESY5 data combinations, as well as the significance levels $N_{\sigma}$ at which these models are preferred over $\Lambda$CDM.}
	\label{tab:2}                        
	\setlength{\tabcolsep}{2mm}
	\renewcommand{\arraystretch}{1.2}
		\footnotesize
			\begin{tabular}{@{\hspace{0.6cm}}l@{\hspace{0.6cm}}c@{\hspace{0.6cm}}c}
				\hline 
				\hline
				Model/Dataset & $\Delta\chi^2_{min}$ & $N_{\sigma}$ \\
				\hline
				
				$\bm{\Lambda w_2\textbf{DM}}$ &  &\\
				CMB+DESI & $5.37 $ &  $2.0\sigma  $ \\
				CMB+DESI+PP & $4.68 $ &  $1.9\sigma  $ \\
				CMB+DESI+Union3 & $4.78 $ & $1.9\sigma  $\\
				CMB+DESI+DESY5 & $4.09   $ & $ 1.7\sigma  $\\
				\hline
				
				$\bm{{w}\textbf{CDM}}$ &  &\\
				CMB+DESI & $0.26 $ &  $-0.3\sigma  $ \\
			    CMB+DESI+PP & $1.22 $ &  $ 0.6\sigma $ \\
			    CMB+DESI+Union3 & $1.14 $ & $ 0.6\sigma $\\
			    CMB+DESI+DESY5 & $ 5.23  $ & $ 2.0\sigma  $\\
			    \hline
			    
			    $\bm{{ww_2}\textbf{DM}}$ &  &\\
			    CMB+DESI & $6.29 $ &  $1.7\sigma  $ \\
			    CMB+DESI+PP & $9.38 $ &  $2.4\sigma  $ \\
			    CMB+DESI+Union3 & $10.87 $ & $2.6\sigma  $\\
			    CMB+DESI+DESY5 & $ 16.41  $ & $ 3.5\sigma  $\\
			    \hline				
				  $\bm{{w_0w_a}\textbf{CDM}}$ &  &\\
				 CMB+DESI & $7.59 $ &  $ 2.0\sigma $ \\
				 CMB+DESI+PP & $7.8 $ &  $2.0\sigma  $ \\
				 CMB+DESI+Union3 & $14.02 $ & $3.1\sigma$\\
				 CMB+DESI+DESY5 & $18.66$ & $ 3.7\sigma $\\
				\hline
				 $\bm{{w_0w_aw_2}\textbf{DM}}$ &  &\\
				CMB+DESI & $8.46 $ &  $ 1.8\sigma $ \\
				CMB+DESI+PP & $10.70 $ &  $ 2.2\sigma $ \\
				CMB+DESI+Union3 & $15.35 $ & $3.0\sigma  $\\
				CMB+DESI+DESY5 & $ 20.39  $ & $3.6\sigma   $\\
				\hline
				\hline
			\end{tabular}
\end{table}

Finally, in Tab.~\ref{tab:2}, we present the minimum chi-square differences of the $\Lambda$CDM model  relative to all other models, defined as $\Delta \chi^2_{\rm min} = \chi^2_{\rm min, \Lambda CDM} - \chi^2_{\rm min}$, as well as the significance levels $N_{\sigma}$ at which these models are preferred over $\Lambda$CDM. The significance is determined by the formula $F_{\chi^2}(\Delta \chi^2_{\rm min};k)=\int_{-\infty}^{N_{\sigma}}\frac{1}{\sqrt{2\pi}}e^{-\frac{x^2}{2}}{\rm d}x$, where $F_{\chi^2}(\Delta \chi^2_{\rm min}; k)$ is the cumulative distribution function of the chi-square distribution with $k$ degrees of freedom, and $k$ is the number of additional free parameters in the given model relative to $\Lambda$CDM. $N_{\sigma}<0$ can be viewed in a non-rigorous way as meaning that the probability of the given models being favored over the $\Lambda$CDM model is less than 50\%. From Tab.~\ref{tab:2}, we find that, for all data combinations, both $ww_2$DM (with $N_{\sigma}$ ranging from $1.7\sigma$ to $3.5\sigma$) and $w_0w_aw_2$DM (with $N_{\sigma}$ ranging from $1.8\sigma$ to $3.6\sigma$) are favored over $\Lambda$CDM, with a significance comparable to that of $w_0w_a$CDM (whose $N_{\sigma}$ ranges from $2.0\sigma$ to $3.7\sigma$).
\section{conclusions}
\label{sec:4}
In this work, we propose a NSDM model in which the EoS of DM is parameterized as $w_{\rm dm}=w_2 a^2$, and then we analyze how this type of DM deviates from CDM under three different DE scenarios, which are the cosmological constant with $w_{\rm de}=-1$, the constant-EoS-parameter DE with $w_{\rm de}=w$, and the CPL DE with $w_{\rm de}=w_0+w_a(1-a)$, using the CMB and DESI BAO datasets, together with three different SN Ia datasets, namely PP, Union3, and DESY5. We also attempt to understand the extent to which the introduction of this NSDM helps to avoid the violation of the null energy condition. The three models formed by combining the proposed NSDM with three types of DE is referred as $\Lambda w_2$DM, $ww_2$DM, and $w_0w_aw_2$DM, respectively.

The analysis results reveal that the three models, $\Lambda w_2$DM, $ww_2$DM, and $w_0w_aw_2$DM, all favor a negative mean value of $w_2$ under the data combinations CMB+DESI, CMB+DESI+PP, CMB+DESI+Union3, and CMB+DESI+DESY5. Moreover, for the $\Lambda w_2$DM model, the preference for a non-zero DM EoS parameter lies in the range of $2.0$-$2.3\sigma$ CL for the considered data combinations. For the $ww_2$DM model, this preference lies in the range of $2.4$-$3.3\sigma$ CL, while for the $w_0w_aw_2$DM model, the preference is below $2\sigma$ CL in all cases. Therefore, it can be seen that considering the CPL DE sector weakens the non-cold behavior of DM compared to the cases with the other two DE models.

By calculating the significance levels $N_{\sigma}$ for the $w$CDM and $ww_2$DM models violating the null energy condition using the formula: $\frac{A_{w<-1}}{A_{\rm tot}}=\int_{-\infty}^{N_{\sigma}}\frac{1}{\sqrt{2\pi}}e^{-\frac{x^2}{2}}{\rm d}x$, where  $A_{w<-1}$ denotes the area under the $w$ distribution where $w < -1$, and $A_{\rm tot}$ denotes the total area under the $w$ distribution. we find that $N_{\sigma}$ decreases from the range of $-2.3\sigma$ to $0.6\sigma$ in the $w$CDM model to the range of $-3.5\sigma$ to $-1.0\sigma$ in the $ww_2$DM model for the considered data combinations.
This indicates that replacing CDM with NSDM in the $w$CDM model can significantly reduce the probability of violating the null energy condition.

Similarly, by calculating the significance levels $N_{\sigma}$ for the $w_0w_a$CDM and $w_0w_aw_2$DM models violating the null energy condition using the formula:$\frac{A_{0<a_{\mathrm{across}}<1}}{A_{\rm tot}}=\int_{-\infty}^{N_{\sigma}}\frac{1}{\sqrt{2\pi}}e^{-\frac{x^2}{2}}{\rm d}x.$ , where  $A_{0<a_{\mathrm{across}}<1}$ denotes the area under the  $a_{\mathrm{across}}$ (the scale factor at which the DE EoS crosses the phantom divide line) distribution where $0<a_{\mathrm{across}}<1$, and $A_{\rm tot}$ denotes the total area under the $a_{\mathrm{across}}$ distribution. we find that $N_{\sigma}$ decreases from the range of $2.7\sigma$ to $4.5\sigma$ in the $w_0w_a$CDM model to the range of $1.2\sigma$ to $2.2\sigma$ in the $w_0w_aw_2$DM model for the considered data combinations.
This indicates that replacing CDM with NSDM in the $w_0w_a$CDM model can significantly reduce the probability of violating the null energy condition.

In the end, by calculating the significance levels $N_{\sigma}$ at which other models considered in this work are preferred over $\Lambda$CDM using the formula:
 $F_{\chi^2}(\Delta \chi^2_{\rm min};k)=\int_{-\infty}^{N_{\sigma}}\frac{1}{\sqrt{2\pi}}e^{-\frac{x^2}{2}}{\rm d}x$, where $F_{\chi^2}(\Delta \chi^2_{\rm min}; k)$ is the cumulative distribution function of the chi-square distribution with $k$ degrees of freedom,  $\Delta \chi^2_{\rm min} = \chi^2_{\rm min, \Lambda CDM} - \chi^2_{\rm min}$ is the minimum chi-square differences of the $\Lambda$CDM model relative to all other models, and $k$ is the number of additional free parameters in the given model relative to $\Lambda$CDM. We find that, for all data combinations, $N_{\sigma}$ in the $ww_2$DM model lies in the range of $1.7\sigma$ to $3.5\sigma$, while that in the $w_0w_aw_2$DM model lies in the range of $1.8\sigma$ to $3.6\sigma$, compared with the $w_0w_a$CDM model, whose $N_{\sigma}$ ranges from $2.0\sigma$ to $3.7\sigma$. This indicates that both $ww_2$DM and $w_0w_aw_2$DM are favored over $\Lambda$CDM with a significance comparable to that of $w_0w_a$CDM.
\section*{Acknowledgments}
This research is supported in part by 
the National Natural Science Foundation of China key project under Grant No. 12535002.
\bibliographystyle{spphys}
\bibliography{w2dm}

\begin{thebibliography}{10}
\providecommand{\url}[1]{{#1}}
\providecommand{\urlprefix}{URL }
\expandafter\ifx\csname urlstyle\endcsname\relax
  \providecommand{\doi}[1]{DOI \discretionary{}{}{}#1}\else
  \providecommand{\doi}{DOI \discretionary{}{}{}\begingroup \urlstyle{rm}\Url}\fi

\bibitem{rubin1970rotation}
V.C. Rubin, W.K. Ford~Jr, Astrophysical Journal, vol. 159, p. 379 \textbf{159}, 379 (1970)

\bibitem{tyson1990detection}
J.A. Tyson, F.~Valdes, R.~Wenk, Astrophysical Journal, Part 2-Letters (ISSN 0004-637X), vol. 349, Jan. 20, 1990, p. L1-L4. \textbf{349}, L1 (1990)

\bibitem{clowe2004weak}
D.~Clowe, A.~Gonzalez, M.~Markevitch, The Astrophysical Journal \textbf{604}(2), 596 (2004)

\bibitem{massey2007dark}
R.~Massey, J.~Rhodes, R.~Ellis, N.~Scoville, A.~Leauthaud, A.~Finoguenov, P.~Capak, D.~Bacon, H.~Aussel, J.P. Kneib, et~al., Nature \textbf{445}(7125), 286 (2007)

\bibitem{smoot1992structure}
G.F. Smoot, C.L. Bennett, A.~Kogut, E.L. Wright, J.~Aymon, N.W. Boggess, E.S. Cheng, G.~De~Amici, S.~Gulkis, M.~Hauser, et~al., Astrophysical Journal, Part 2-Letters (ISSN 0004-637X), vol. 396, no. 1, Sept. 1, 1992, p. L1-L5. Research supported by NASA. \textbf{396}, L1 (1992)

\bibitem{spergel2003first}
D.N. Spergel, L.~Verde, H.V. Peiris, E.~Komatsu, M.~Nolta, C.L. Bennett, M.~Halpern, G.~Hinshaw, N.~Jarosik, A.~Kogut, et~al., The Astrophysical Journal Supplement Series \textbf{148}(1), 175 (2003)

\bibitem{spergel2007three}
D.N. Spergel, R.~Bean, O.~Dor{\'e}, M.~Nolta, C.~Bennett, J.~Dunkley, G.~Hinshaw, N.e. Jarosik, E.~Komatsu, L.~Page, et~al., The astrophysical journal supplement series \textbf{170}(2), 377 (2007)

\bibitem{hinshaw2013nine}
G.~Hinshaw, D.~Larson, E.~Komatsu, D.N. Spergel, C.~Bennett, J.~Dunkley, M.~Nolta, M.~Halpern, R.~Hill, N.~Odegard, et~al., The Astrophysical Journal Supplement Series \textbf{208}(2), 19 (2013)

\bibitem{ade2014planck}
P.A. Ade, N.~Aghanim, C.~Armitage-Caplan, M.~Arnaud, M.~Ashdown, F.~Atrio-Barandela, J.~Aumont, C.~Baccigalupi, A.J. Banday, R.~Barreiro, et~al., Astronomy \& Astrophysics \textbf{571}, A16 (2014)

\bibitem{planck2016planck}
P.~Collaboration, et~al., Astronomy and Astrophysics \textbf{594}, A13 (2016)

\bibitem{aghanim2020planck-1}
N.~Aghanim, Y.~Akrami, M.~Ashdown, J.~Aumont, C.~Baccigalupi, M.~Ballardini, A.J. Banday, R.~Barreiro, N.~Bartolo, S.~Basak, et~al., Astronomy \& Astrophysics \textbf{641}, A6 (2020)

\bibitem{davis1985evolution}
M.~Davis, G.~Efstathiou, C.S. Frenk, S.D. White, Astrophysical Journal, Part 1 (ISSN 0004-637X), vol. 292, May 15, 1985, p. 371-394. Research supported by the Science and Engineering Research Council of England and NASA. \textbf{292}, 371 (1985)

\bibitem{baugh1993three}
C.~Baugh, G.~Efstathiou, Monthly Notices of the Royal Astronomical Society \textbf{265}(1), 145 (1993)

\bibitem{colless20012df}
M.~Colless, G.~Dalton, S.~Maddox, W.~Sutherland, P.~Norberg, S.~Cole, J.~Bland-Hawthorn, T.~Bridges, R.~Cannon, C.~Collins, et~al., Monthly Notices of the Royal Astronomical Society \textbf{328}(4), 1039 (2001)

\bibitem{percival20012df}
W.J. Percival, C.M. Baugh, J.~Bland-Hawthorn, T.~Bridges, R.~Cannon, S.~Cole, M.~Colless, C.~Collins, W.~Couch, G.~Dalton, et~al., Monthly Notices of the Royal Astronomical Society \textbf{327}(4), 1297 (2001)

\bibitem{tegmark2004three}
M.~Tegmark, M.R. Blanton, M.A. Strauss, F.~Hoyle, D.~Schlegel, R.~Scoccimarro, M.S. Vogeley, D.H. Weinberg, I.~Zehavi, A.~Berlind, et~al., The Astrophysical Journal \textbf{606}(2), 702 (2004)

\bibitem{eisenstein2005detection}
D.J. Eisenstein, I.~Zehavi, D.W. Hogg, R.~Scoccimarro, M.R. Blanton, R.C. Nichol, R.~Scranton, H.J. Seo, M.~Tegmark, Z.~Zheng, et~al., The Astrophysical Journal \textbf{633}(2), 560 (2005)

\bibitem{riess1998observational}
A.G. Riess, A.V. Filippenko, P.~Challis, A.~Clocchiatti, A.~Diercks, P.M. Garnavich, R.L. Gilliland, C.J. Hogan, S.~Jha, R.P. Kirshner, et~al., The Astronomical Journal \textbf{116}(3), 1009 (1998)

\bibitem{perlmutter1999measurements}
S.~Perlmutter, G.~Aldering, G.~Goldhaber, R.~Knop, P.~Nugent, P.~Castro, S.~Deustua, S.~Fabbro, A.~Goobar, D.~Groom, et~al., The Astrophysical Journal \textbf{517}(2), 565 (1999)

\bibitem{anderson2014clustering}
L.~Anderson, E.~Aubourg, S.~Bailey, F.~Beutler, V.~Bhardwaj, M.~Blanton, A.S. Bolton, J.~Brinkmann, J.R. Brownstein, A.~Burden, et~al., Monthly Notices of the Royal Astronomical Society \textbf{441}(1), 24 (2014)

\bibitem{alam2017clustering}
S.~Alam, M.~Ata, S.~Bailey, F.~Beutler, D.~Bizyaev, J.A. Blazek, A.S. Bolton, J.R. Brownstein, A.~Burden, C.H. Chuang, et~al., Monthly Notices of the Royal Astronomical Society \textbf{470}(3), 2617 (2017)

\bibitem{alam2021completed}
S.~Alam, M.~Aubert, S.~Avila, C.~Balland, J.E. Bautista, M.A. Bershady, D.~Bizyaev, M.R. Blanton, A.S. Bolton, J.~Bovy, et~al., Physical Review D \textbf{103}(8), 083533 (2021)

\bibitem{klypin1999missing}
A.~Klypin, A.V. Kravtsov, O.~Valenzuela, F.~Prada, The Astrophysical Journal \textbf{522}(1), 82 (1999)

\bibitem{moore1999dark}
B.~Moore, S.~Ghigna, F.~Governato, G.~Lake, T.~Quinn, J.~Stadel, P.~Tozzi, The Astrophysical Journal \textbf{524}(1), L19 (1999)

\bibitem{boylan2012milky}
M.~Boylan-Kolchin, J.S. Bullock, M.~Kaplinghat, Monthly Notices of the Royal Astronomical Society \textbf{422}(2), 1203 (2012)

\bibitem{moore1999cold}
B.~Moore, T.~Quinn, F.~Governato, J.~Stadel, G.~Lake, Monthly Notices of the Royal Astronomical Society \textbf{310}(4), 1147 (1999)

\bibitem{springel2008aquarius}
V.~Springel, J.~Wang, M.~Vogelsberger, A.~Ludlow, A.~Jenkins, A.~Helmi, J.F. Navarro, C.S. Frenk, S.D. White, Monthly Notices of the Royal Astronomical Society \textbf{391}(4), 1685 (2008)

\bibitem{blumenthal1982galaxy}
G.R. Blumenthal, H.~Pagels, J.R. Primack, Nature \textbf{299}(5878), 37 (1982)

\bibitem{bode2001halo}
P.~Bode, J.P. Ostriker, N.~Turok, The Astrophysical Journal \textbf{556}(1), 93 (2001)

\bibitem{hu2000fuzzy}
W.~Hu, R.~Barkana, A.~Gruzinov, Physical Review Letters \textbf{85}(6), 1158 (2000)

\bibitem{marsh2014model}
D.J. Marsh, J.~Silk, Monthly Notices of the Royal Astronomical Society \textbf{437}(3), 2652 (2014)

\bibitem{spergel2000observational}
D.N. Spergel, P.J. Steinhardt, Physical review letters \textbf{84}(17), 3760 (2000)

\bibitem{wang2014cosmological}
M.Y. Wang, A.H. Peter, L.E. Strigari, A.R. Zentner, B.~Arant, S.~Garrison-Kimmel, M.~Rocha, Monthly Notices of the Royal Astronomical Society \textbf{445}(1), 614 (2014)

\bibitem{mueller2005cosmological}
C.M. Mueller, Physical Review D \textbf{71}(4), 047302 (2005)

\bibitem{Kumar:2012gr}
S.~Kumar, L.~Xu, Phys. Lett. B \textbf{737}, 244 (2014)

\bibitem{kumar2014observational}
S.~Kumar, L.~Xu, Physics Letters B \textbf{737}, 244 (2014)

\bibitem{Murgia:2017lwo}
R.~Murgia, A.~Merle, M.~Viel, M.~Totzauer, A.~Schneider, JCAP \textbf{11}, 046 (2017)

\bibitem{Gariazzo:2017pzb}
S.~Gariazzo, M.~Escudero, R.~Diamanti, O.~Mena, Phys. Rev. D \textbf{96}(4), 043501 (2017)

\bibitem{Murgia:2018now}
R.~Murgia, V.~Ir\v{s}i\v{c}, M.~Viel, Phys. Rev. D \textbf{98}(8), 083540 (2018)

\bibitem{kopp2018dark}
M.~Kopp, C.~Skordis, D.B. Thomas, S.~Ili{\'c}, Physical Review Letters \textbf{120}(22), 221102 (2018)

\bibitem{Schneider:2018xba}
A.~Schneider, Phys. Rev. D \textbf{98}(6), 063021 (2018)

\bibitem{kumar2019testing}
S.~Kumar, R.C. Nunes, S.K. Yadav, Monthly Notices of the Royal Astronomical Society \textbf{490}(1), 1406 (2019)

\bibitem{ilic2021dark}
S.~Ili{\'c}, M.~Kopp, C.~Skordis, D.B. Thomas, Physical Review D \textbf{104}(4), 043520 (2021)

\bibitem{Najera:2020smt}
S.~N\'ajera, R.A. Sussman, Eur. Phys. J. C \textbf{81}(4), 374 (2021)

\bibitem{pan2023iwdm}
S.~Pan, W.~Yang, E.~Di~Valentino, D.F. Mota, J.~Silk, Journal of Cosmology and Astroparticle Physics \textbf{2023}(07), 064 (2023)

\bibitem{yao2024observational}
Y.H. Yao, J.C. Wang, X.H. Meng, Physical Review D \textbf{109}(6), 063502 (2024)

\bibitem{Yao:2023qve}
Y.H. Yao, X.H. Meng, Commun. Theor. Phys. \textbf{76}(7), 075401 (2024)

\bibitem{Wang:2025zri}
D.~Wang,   (2025)

\bibitem{Kumar:2025etf}
U.~Kumar, A.~Ajith, A.~Verma,   (2025)

\bibitem{Yang:2025ume}
W.~Yang, S.~Pan, E.~Di~Valentino, O.~Mena, D.F. Mota, S.~Chakraborty, Phys. Rev. D \textbf{111}(10), 103509 (2025)

\bibitem{Liu:2025mob}
J.Q. Liu, Y.H. Yao, Y.~Su, J.W. Wu, Res. Astron. Astrophys. \textbf{25}(7) (2025)

\bibitem{Yao:2025kuz}
Y.H. Yao, J.Q. Liu, Phys. Dark Univ. \textbf{49}, 102052 (2025)

\bibitem{Li:2025eqh}
T.N. Li, Y.M. Zhang, Y.H. Yao, P.J. Wu, J.F. Zhang, X.~Zhang,   (2025)

\bibitem{Li:2025dwz}
T.N. Li, P.J. Wu, G.H. Du, Y.H. Yao, J.F. Zhang, X.~Zhang, Phys. Dark Univ. \textbf{50}, 102068 (2025)

\bibitem{DESI:2025zgx}
M.~Abdul~Karim, et~al.,   (2025)

\bibitem{Li:2024qso}
T.N. Li, P.J. Wu, G.H. Du, S.J. Jin, H.L. Li, J.F. Zhang, X.~Zhang, Astrophys. J. \textbf{976}(1), 1 (2024)

\bibitem{Giare:2024gpk}
W.~Giar\`e, M.~Najafi, S.~Pan, E.~Di~Valentino, J.T. Firouzjaee, JCAP \textbf{10}, 035 (2024)

\bibitem{Dinda:2024ktd}
B.R. Dinda, R.~Maartens, JCAP \textbf{01}, 120 (2025)

\bibitem{Escamilla:2024ahl}
L.A. Escamilla, E.~\"Oz\"ulker, O.~Akarsu, E.~Di~Valentino, J.A. V\'azquez,   (2024)

\bibitem{Sabogal:2024yha}
M.A. Sabogal, E.~Silva, R.C. Nunes, S.~Kumar, E.~Di~Valentino, W.~Giar\`e, Phys. Rev. D \textbf{110}(12), 123508 (2024)

\bibitem{Li:2024qus}
T.N. Li, Y.H. Li, G.H. Du, P.J. Wu, L.~Feng, J.F. Zhang, X.~Zhang, Eur. Phys. J. C \textbf{85}(6), 608 (2025)

\bibitem{Li:2024hrv}
J.X. Li, S.~Wang,   (2024)

\bibitem{Wang:2024dka}
H.~Wang, Y.S. Piao,   (2024)

\bibitem{Huang:2025som}
L.~Huang, R.G. Cai, S.J. Wang,   (2025)

\bibitem{Li:2025owk}
T.N. Li, G.H. Du, Y.H. Li, P.J. Wu, S.J. Jin, J.F. Zhang, X.~Zhang,   (2025)

\bibitem{Wu:2025vfs}
P.J. Wu, T.N. Li, G.H. Du, X.~Zhang,   (2025)

\bibitem{Li:2025ula}
Y.H. Li, X.~Zhang,   (2025)

\bibitem{Li:2025ops}
J.X. Li, S.~Wang,   (2025)

\bibitem{Barua:2025ypw}
S.~Barua, S.~Desai,   (2025)

\bibitem{Yashiki:2025loj}
M.~Yashiki,   (2025)

\bibitem{Ling:2025lmw}
J.L. Ling, G.H. Du, T.N. Li, J.F. Zhang, S.J. Wang, X.~Zhang,   (2025)

\bibitem{Goswami:2025uih}
S.~Goswami, S.~Das, Phys. Dark Univ. \textbf{48}, 101951 (2025)

\bibitem{Yang:2025boq}
Y.~Yang, X.~Dai, Y.~Wang, Phys. Rev. D \textbf{111}(10), 103534 (2025)

\bibitem{Pang:2025lvh}
Y.H. Pang, X.~Zhang, Q.G. Huang, Sci. China Phys. Mech. Astron. \textbf{68}(8), 280410 (2025)

\bibitem{You:2025uon}
C.~You, D.~Wang, T.~Yang,   (2025)

\bibitem{Ozulker:2025ehg}
E.~{\"O}z{\"u}lker, E.~Di~Valentino, W.~Giar{\`e},   (2025)

\bibitem{Cheng:2025lod}
H.~Cheng, E.~Di~Valentino, L.A. Escamilla, A.A. Sen, L.~Visinelli,   (2025)

\bibitem{Pan:2025qwy}
S.~Pan, S.~Paul, E.N. Saridakis, W.~Yang,   (2025)

\bibitem{Li:2025muv}
T.N. Li, G.H. Du, Y.H. Li, Y.~Li, J.L. Ling, J.F. Zhang, X.~Zhang,   (2025)

\bibitem{Du:2025xes}
G.H. Du, T.N. Li, P.J. Wu, J.F. Zhang, X.~Zhang,   (2025)

\bibitem{Chen:2025wwn}
X.~Chen, A.~Loeb,   (2025)

\bibitem{Giani:2025hhs}
L.~Giani, R.~Von~Marttens, O.F. Piattella,   (2025)

\bibitem{Braglia:2025gdo}
M.~Braglia, X.~Chen, A.~Loeb,   (2025)

\bibitem{Wang:2024rus}
J.~Wang, Z.~Huang, Y.~Yao, J.~Liu, L.~Huang, Y.~Su, JCAP \textbf{09}, 053 (2024).
\newblock [Erratum: JCAP 09, E01 (2025)]

\bibitem{Su:2025ntt}
Y.~Su, Z.~Huang, J.~Wang, Y.~Yao, J.~Liu,   (2025)

\bibitem{Kou:2025yfr}
R.~Kou, A.~Lewis,   (2025)

\bibitem{Rubin:2023jdq}
D.~Rubin, et~al.,   (2023)

\bibitem{hu1996small}
W.~Hu, N.~Sugiyama, The Astrophysical Journal \textbf{471}(2), 542 (1996)

\bibitem{DES:2024jxu}
T.M.C. Abbott, et~al., Astrophys. J. Lett. \textbf{973}(1), L14 (2024)

\bibitem{scolnic2022pantheon+}
D.~Scolnic, D.~Brout, A.~Carr, A.G. Riess, T.M. Davis, A.~Dwomoh, D.O. Jones, N.~Ali, P.~Charvu, R.~Chen, et~al., The Astrophysical Journal \textbf{938}(2), 113 (2022)

\bibitem{foreman2013emcee}
D.~Foreman-Mackey, D.W. Hogg, D.~Lang, J.~Goodman, Publications of the Astronomical Society of the Pacific \textbf{125}(925), 306 (2013)

\bibitem{Lewis:2019xzd}
A.~Lewis,   (2019).
\newblock \urlprefix\url{https://getdist.readthedocs.io}

\end{thebibliography}

\end{document}